# Miniaturization of the Superconducting Memory Cell *via* a Three-Dimensional Nb Nano-Superconducting Quantum Interference Device

Lei Chen[1,2,*], Lili Wu[1,2], Yue Wang[1,2], Yinping Pan[1,4], Denghui Zhang[1,2], Junwen Zeng[1], Xiaoyu Liu[1], Linxian Ma[1], Wei Peng[1,2], Yihua Wang[4], Jie Ren[1,2], Zhen Wang[1,2,3,*]

[1]CAS Center for Excellence in Superconducting Electronics (CENSE), State Key Laboratory of Functional Material for Informatics, Shanghai Institute of Microsystem and Information Technology (SIMIT), Chinese Academy of Sciences (CAS), Shanghai 200050, China

[2]University of the Chinese Academy of Sciences, Beijing 100049, China

[3]School of Physical Science and Technology, Shanghai Tech University, Shanghai 200031, China

[4]Department of Physics and State Key Laboratory of Surface Physics, Fudan University, Shanghai 200438 China

*Corresponding authors: leichen@mail.sim.ac.cn; zwang@mail.sim.ac.cn

**Abstract:** Scalable memories that can match the speeds of superconducting logic circuits have long been desired to enable a superconducting computer. A superconducting loop that includes a Josephson junction can store a flux quantum state in picoseconds. However, the requirement for the loop inductance to create a bi-state hysteresis sets a limit on the minimal area occupied by a single memory cell. Here, we present a miniaturized superconducting memory cell based on a Three-Dimensional (3D) Nb nano-Superconducting QUantum Interference Device (nano-SQUID). The major cell area here fits within an 8×9 μm$^2$ rectangle with a cross-selected function for memory implementation. The cell shows periodic tunable hysteresis between two neighbouring flux quantum states produced by bias current sweeping because of the large modulation depth of the 3D nano-SQUID (~66%). Furthermore, the measured Current-Phase Relations (CPRs) of nano-SQUIDs are shown to be skewed from a sine function, as predicted by theoretical modelling. The skewness and the critical current of 3D nano-SQUIDs are linearly correlated. It is also found that the hysteresis loop size is in a linear scaling relationship with the CPR skewness using the statistics from characterisation of 26 devices. We show that the CPR skewness range of $\pi/4$–$3\pi/4$ is equivalent to a large loop inductance in creating a stable bi-state hysteresis for memory implementation. Therefore, the skewed CPR of 3D nano-SQUID enables further superconducting memory cell miniaturization by overcoming the inductance limitation of the loop area.

As computers based on complementary metal-oxide-semiconductor (CMOS) technology approach the limits of their device physics, superconducting digital circuits based on Single-Flux-Quantum (SFQ) with picosecond speed and attowatt power consumption are becoming increasingly attractive.[1-6] Their cryogenic working temperatures also make them naturally compatible with building high-frequency controlling circuits for future quantum computers.[7,8] However, the lack of a compatible high-density, high-speed memory has been a long-standing bottleneck.[1,9,10] Current SFQ cache memories are mainly composed of volatile shift-register units.[11,12] The storage element area is approximately 60×60 μm$^2$. This size not only limits the memory capacity on a single chip but also its speed because signal propagation times among cells become nonnegligible in the picosecond regime.

The hybrid memories that combine CMOS memories with SFQ input/output adapter[13,14] are the most straightforward technical solution when the integration scalability of CMOS manufacturing lines is taken into account, despite their high-power consumption and speed limitations. Recent developments in superconducting spintronics involving combination of the magnetism and the superconductivity at the junction level have mitigated the power consumption problem.[15-20] However, the magnetic polarisation switching speed is restrained by the magnetic junction physics at cryogenic temperatures. Use of flux quantum in a superconducting loop to store data can match the speeds of SFQ circuits.[21] Several types of memory cells like Vortex Transition Memory and Quantum Flux Parametron Memory have been developed by tunnelling Josephson junctions.[22-24] Recently, these cell dimensions were reduced to 9×11 μm$^2$ by using a state-of-the-art Nb-based-junction process developed at the MIT Lincoln Laboratory.[10] Unfortunately, further cell area reduction will be extremely challenging because of the requirement for few-pH inductors. Therefore, the problem is being tackled *via* other non-traditional methods including superconducting nanowires[25,26] or planar nano-superconducting quantum interference devices (nano-SQUIDs)[27] with large kinetic inductance that can replace geometric inductances. However, these devices with a linear current-phase relation (CPR) did not show a stable bi-state hysteresis similar to that of ferromagnets for the storage of binary data.

Nano-SQUIDs of Dayem-type nanobridge junctions (NBJs) have been developed to study the nanoscale magnetism for many years.[28-34] The 3D Nb nano-SQUID that we developed recently has demonstrated a decent Josephson effect and a relatively large kinetic inductance.[35,36] According to theoretical models, the CPR of the NBJ is skewed from that of the ideal Josephson effect.[37-40] The shape of CPR plays an important role in the storage hysteresis between flux quantum states. Recently, the CPRs of several non-traditional Josephson junctions have been studied to reveal their intrinsic physical properties.[41-43] However, the exact CPRs of NBJs have not yet been fully studied experimentally.[44]

Here, we developed a superconducting memory cell using a 3D nano-SQUID. We demonstrated a hysteresis loop in flux quantum states by sweeping the biasing current to the storage loop. The hysteresis loop size can be further tuned by flux tuning of the nano-SQUID for cross-selected memory cell implementation. Additionally, the CPR of the nano-SQUID and the single NBJs in the storage loop was obtained from the bias current as a function of the total magnetic flux. The measured CPRs show good consistency with the theoretical model and are skewed from a sine function. Furthermore, we found that the CPR skewness of 3D nano-SQUIDs plays an equivalent role to the loop inductance in forming a stable memory hysteresis loop. We concluded that a memory cell based on a 3D nano-SQUID with a skewed CPR can lift the inductance constraints and thus is very promising for scalable superconducting memory.

**Results and discussion**

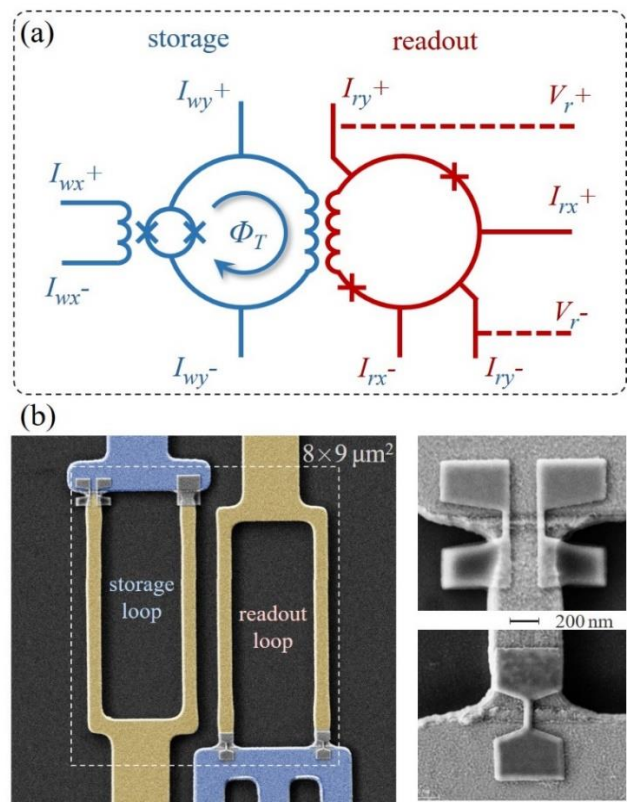

**Figure 1.** (a) Schematic diagram of the memory cell. (b) False-coloured scanning electron microscope (SEM) images of the memory cell. The inset on the right shows magnified images of the 3D nano-SQUID and the single nanobridge junction (NBJ).

Figure 1(a) shows a schematic diagram of the miniaturized memory cell. The memory cell consists of a storage loop (blue) and a readout micro-SQUID (red). The storage loop includes a 3D nano-SQUID that acts as a flux-tunable Josephson junction. The datum "0" or "1" value is written into the storage loop as a flux quantum state $\Phi_T$ by ramping the bias current $I_{wy}$. The external coil current $I_{wx}$ can select the written data condition of the storage loop by flux modulation of the critical current of the nano-SQUID. The storage loop is inductively coupled to the readout micro-SQUID so that the reading and writing operations are independent from each other and the data can be read out non-destructively. The micro-SQUID used here is also made from 3D nanobridge junctions. The readout micro-SQUID is locked at an operating point by selecting a bias current $I_{rx}$ and a flux feedback current $I_{ry}$. When the flux quantum state $\Phi_T$ in the storage loop jumps, a voltage change across the SQUID will be sensed as a data readout. Both the storage loop and readout micro-SQUIDs can be cross-selected in an XY manner for memory implementation of both the data and the address. A false-coloured SEM image of the complete memory cell is shown in Figure 1(b). The blue and yellow structures are made from 150-nm-thick Nb films and are electrically separated by a 12-nm-wide $SiO_2$ slit. Then, a 10-nm-thick and 50-nm-wide Nb nanowire are set across the insulated slit to form a 3D NBJ. The insets on the right show magnified scanning electron microscope (SEM) images of a nano-SQUID and a single NBJ.

To write a binary datum, the storage loop should switch steadily between two flux quantum states in a manner similar to ferromagnetic hysteresis. In Figure 2(a) and (b), the variation of the magnetic flux trapped in the storage loop $\Phi_T$ is measured using the readout micro-SQUID in terms of the feedback current $I_{rx}$ in the flux-locking mode. The storage loop bias current $I_{wy}$ was swept in both the positive (blue) and negative (red) directions. Figure 2(a) and (b) correspond to the devices composed of a storage loop with a single NBJ and a nano-SQUID, respectively. In both figures, we observed periodic jumps in $I_{rx}$ that indicate consecutive switching of $\Phi_T$ among several flux quantum states. The period $I_{rx\text{-}period}$ corresponds to a flux quantum $\Phi_0$ in the storage loop. The period $I_{wy\text{-}period}$ along the $I_{wy}$ axis is equal to $\Phi_0/L_r$, where $L_r$ is the inductance of the right part of the storage loop that does not contain the junction. Depending on the sweep direction, a clear hysteresis occurs in $I_{wy}$ between the jumps of two consecutive flux quantum states. Therefore, the data can be stored in different flux quantum states under the same $I_{wy}$ bias by changing the $I_{wy}$ ramping direction. The difference between the $I_{wy}$ values at two jumps is defined as the size of the hysteresis loop $\Delta I_{wy}$, and the jump height is defined as $\Delta\Phi/\Phi_0 = \Delta I_{rx}/I_{rx\text{-}period}$. The insets on the right of Figure 2(a) and (b) show magnified regions near the first jumps between the "0" and "1" flux quanta. Apart from the exact values of $\Delta I_{wy}$ and $\Delta\Phi$, the similarity between the hysteresis profiles of (a) and (b) indicates that the nano-SQUID can be regarded as a single junction under zero magnetic flux bias. Furthermore, we obtained the CPR denoted by $f(\theta)$ for a single junction and a nano-SQUID using the equation $I_0 L_T f(\theta)/\Phi_0 = I_{wy}/I_{wy\text{-}period} - I_{rx}/I_{rx\text{-}period}$ (see the Methods section), as shown in Figure 2(c) and (d), where $L_T$ is the total storage loop inductance. The observed CPRs denoted by $f(\theta)$ are skewed from a standard sine function, as predicted by the theoretical model for the NBJs. Here, we define the skewness $\Delta\theta$ as the phase difference between the maximum point of the CPR and $\pi/2$, which is the maximum point of a sine function.

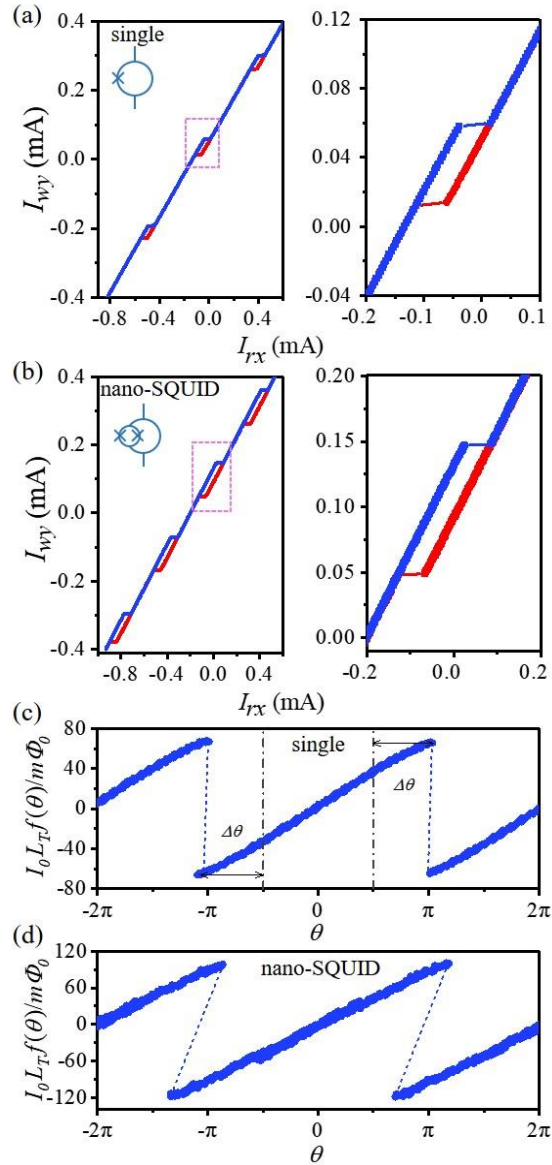

**Figure 2** Trapped magnetic flux measured using the micro-SQUID feedback current $I_{rx}$ as a function of the bias current $I_{wy}$ of the storage loops with (a) a single nanobridge junction and (b) a nano-SQUID. The insets on the right show magnified views of the region of the first jump between flux quantum states. The blue and red colours indicate the positive and negative $I_{wy}$ sweeping directions. (c) and (d) Deduced current-phase relationships of a single junction and a nano-SQUID that correspond to (a) and (b), respectively.

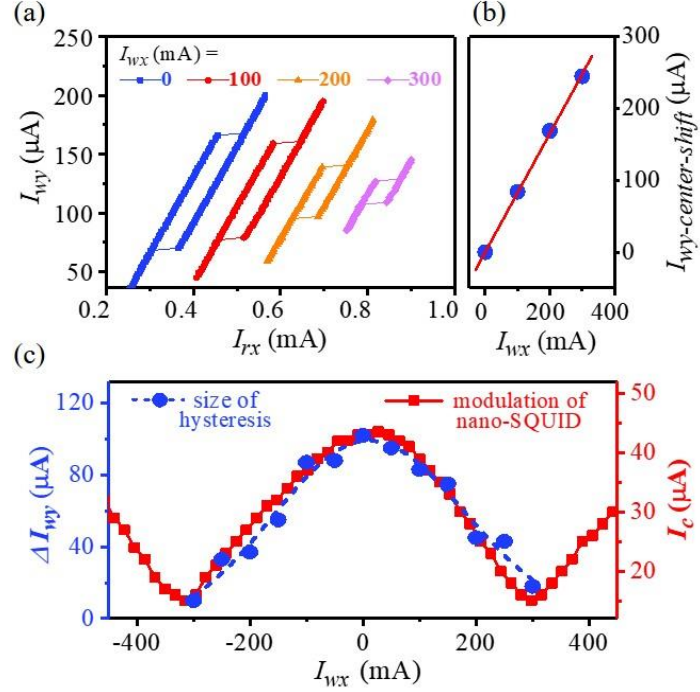

**Figure 3** (a) Hysteresis in $I_{wy}$ of the storage loop acquired with a nano-SQUID at various $I_{wx}$ values through the external coil. (b) Shifts in the centre of hysteresis in (a) along the $I_{wy}$ axis to be aligned on the same vertical level. (c) Size of the hysteresis current $I_{wy}$ for different field coils corresponding to an at-zero magnetic field.

For the cross-selected memory cell implementation, the hysteresis loop size should also be independently tunable. In Figure 3(a), we show the storage loop hysteresis measured at various magnetic flux bias values applied to the nano-SQUID by an external coil with current $I_{wx}$. As $I_{wx}$ increased from 0 to 300 mA, the size of the hysteresis $\Delta I_{wy}$ decreased continuously. The centre of the hysteresis loop was shifted to the same $I_{wy}$ level by the amount $I_{wy\text{-}center\text{-}shift}$ for a better comparison and $I_{wy\text{-}center\text{-}shift}$ was then plotted as a function of $I_{wx}$ in Figure 3(b). The linear fitting of $I_{wy\text{-}center\text{-}shift}$ in Figure 3(b) indicates that the magnetic flux that is coupled to the storage loop acts as a flux bias with a mutual inductance of 0.34 pH. In Figure 3(c), we plotted $\Delta I_{wy}$ as a function of $I_{wx}$ on the left axis (blue) and as a function of the flux modulation of a nano-SQUID on the right axis (red). The tuning of $\Delta I_{wy}$ follows the flux modulation profile of the nano-SQUID. The nano-SQUID has the same dimensions as the nano-SQUID in the storage loop and was measured independently using the same external coil $I_{wx}$. The modulation depth of the 3D nano-SQUID is ~66% and is defined as the percentage of the modulated amplitude with the maximal value of the critical current. The remaining 34% of the unmodulated critical current is usually explained using the skewed CPR model. No change in the size of the hysteresis loop was observed for the storage loop with the single NBJ. This indicates that the magnetic flux that is coupled to the storage loop will not affect its hysteresis. Therefore, the tuning of the hysteresis loop size is caused by the flux modulation of the nano-SQUID alone. The action of the nano-SQUID as a flux-tunable junction thus enables selection of the required memory cell.

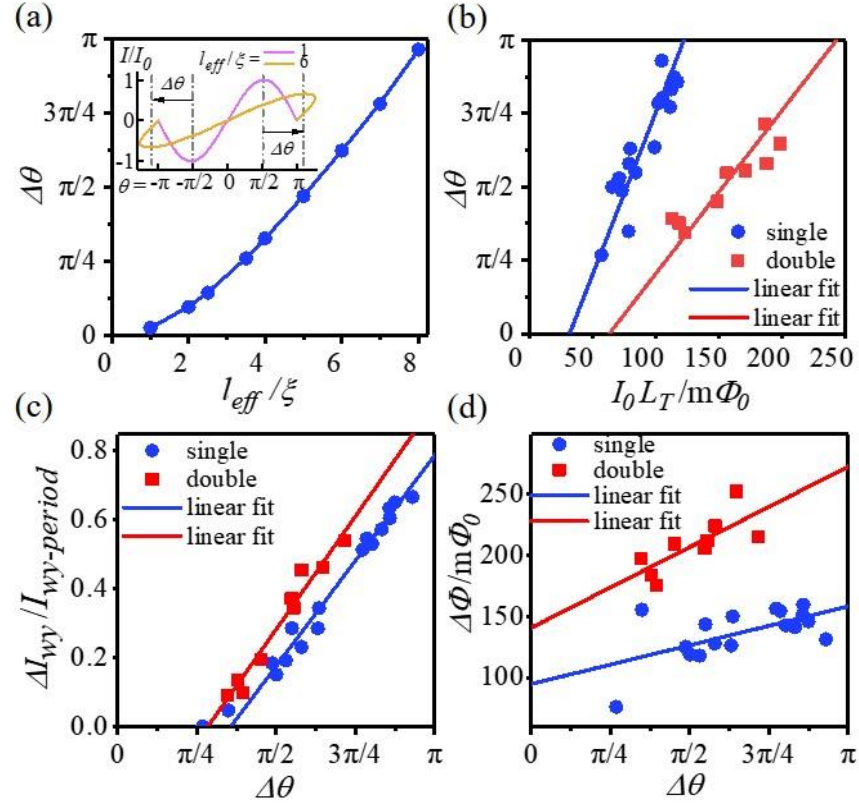

**Figure 4.** (a) CPR skewness $\Delta\theta$ as a function of the ratio of the nanobridge junction effective length to the superconducting length ($l_{eff}/\xi$); the inset shows plots of the calculated CPR for $l_{eff}/\xi=1$ and 6. (b) Plots of $\Delta\theta$ as a function of the CPR amplitude $I_0L_T$ for single junctions and nano-SQUIDs. (c) Plots of the normalised hysteresis loop size $\Delta I_{wy}/I_{wy\text{-}period}$ as a function of $\Delta\theta$. (d) Plots of the flux jump height $\Delta\Phi_T$ as a function of $\Delta\theta$. The blue and red lines in (b), (c) and (d) represent fitting lines for devices formed with a single junction and with a nano-SQUID, respectively.

A nano-SQUID with nanobridge junctions usually has a smaller magnetic flux modulation when compared with that of a traditional SQUID made from tunnelling junctions. A theoretical model based on the Ginzburg-Landau equation predicts that the CPR of a NBJ deviates from the sine function based on the ratio between its effective length $l_{eff}$ [37] and the superconducting coherence length $\xi$.[38, 39, 45] Here, we define the skewness $\Delta\theta$ as the phase difference of the maximum CPR value from $\pi/2$, which is the maximum value of a sine function. In Figure 4(a), we plotted $\Delta\theta$ as a function of $l_{eff}/\xi$. The inset plot shows the calculated CPRs for $l_{eff}/\xi=1$ and $l_{eff}/\xi=6$. When $l_{eff}/\xi=1$, the CPR is an approximate sine wave and $\Delta\theta = 0$; $\Delta\theta$ then starts to increase as $l_{eff}/\xi$ increases. The measured CPRs shown in Figure 2(c) and (d) and Figure 5(b) confirm their consistency with the predictions of the theoretical model and give information on $\Delta\theta$ for each device. We then characterised 17 devices with single NBJs and nine devices with nano-SQUIDs in their storage loops. The 3D NBJs that we developed have improved $\Delta\theta$ considerably when compared with the use of planar nanobridge junctions. However, the CPR skewness $\Delta\theta$ of these junctions still spreads from $\pi/4$ to $\pi$, where the uncertainty comes mainly from the nonuniform junction thickness caused by the lift-off step during the fabrication process. The thickness affects the final $l_{eff}$ value of the junction. In Figure 4(b), we plotted $\Delta\theta$ as a function of the amplitude of the measured CPR $I_0L_T$. The behaviours of both the single junctions and the nano-SQUID scale linearly with $\Delta\theta$. This occurs because the critical current $I_0$ and the effective length $l_{eff}$ are both mainly determined by the junction thickness here. The scaling slopes given by $I_0L_T/\Delta\theta$ of the devices with the single junction and the nano-SQUIDs are 28.8 m$\Phi_0$/rad and 57.2 m$\Phi_0$/rad, respectively. The slope of the device with the nano-SQUID is nearly doubled when compared with that of the single-NBJ device because it consists of two junctions in parallel. In Figure 4(c), the normalised hysteresis loop size $\Delta I_{wy}/I_{wy\text{-}period}$ is also plotted as a function of $\Delta\theta$. Straight lines composed of $\Delta I_{wy}/I_{wy\text{-}period} = 0.39\Delta\theta-0.44$ and $\Delta I_{wy}/I_{wy\text{-}period} = 0.42\Delta\theta-0.37$ can be used to fit the characteristics of the single NBJ and nano-SQUID devices, respectively. The fitting lines show good agreement with our estimated hysteresis loop size $\Delta I_w/I_{w\text{-}period}=2I_0L_T/\Phi_0+(\Delta\theta/\pi)-1/2$. Because the nano-SQUID $I_0$ is double that of the single NBJ, its fitting line is slightly above that for the single NBJ device. Therefore, $\Delta\theta$ is equivalent mathematically to the storage loop inductance $L_T$ in creating a stable memory hysteresis. Tuning of $\Delta\theta$ to $\sim\pi/2$ will mean that it is no longer necessary to have a minimal $L_T$ requirement. It is also important that $\Delta I_{wy}/I_{wy\text{-}period} < 1$ with $\Delta\theta < \pi$ because overlapping of the hysteresis with the next flux quantum states must be avoided to permit definitive state writing. In addition, high skewness in the CPR will also lower the flux modulation depth of the nano-SQUID and lead to an untunable hysteresis size. Therefore, a $\Delta\theta$ range from $\pi/4$ to $3\pi/4$ with a $\Delta I_{wy}/I_{wy\text{-}period}$ ranging from 0 to 0.6 is suitable for an experimentally tunable memory hysteresis. The jump height in the change in the flux quantum states can be estimated using $\Delta\Phi_T = 2I_oL_T/(1+\Delta I_w/I_{w\text{-}period})$. This parameter is a rational function of $\Delta\theta$ and will become saturated at $I_oL_T$. Within the scope of the plot, it can be approximated using a line and provides a guide for the eye to indicate the increasing trend. The slopes here are 20.1 m$\Phi_0$/rad and 41.8 m$\Phi_0$/rad for the devices with the single junctions and the nano-SQUIDs, respectively. These results indicate that a change in $\Delta\theta$ from $\pi/4$ to $3\pi/4$ will only increase $\Delta\Phi_T$ by 31.6 m$\Phi_0$ and 65.7 m$\Phi_0$ for the two devices, respectively, and will not affect the micro-SQUID readout greatly. Therefore, $\Delta\theta$ can play the same role as the storage loop inductance $L_T$ in creating a stable memory hysteresis. The change in $\Delta\theta$ of the 3D nano-SQUID from $\pi/4$ to $3\pi/4$ increases the value of $\Delta I_{wy}/I_{wy\text{-}period}$ from 0 to 0.6, which is good for memory implementation. The jump height of $\Delta\Phi$=206.4±32.9 m$\Phi_0$ at $\Delta\theta=\pi/2\pm\pi/4$ is much larger than the regular resolution of a micro-SQUID, which is less than 1 m$\Phi_0$.

By integrating $I_{wx}$ line onto the chip to modulate of the nano-SQUID,[46-48] the memory cell would be ready for the array implementation. The dimensions of the memory cell developed here are 8×9 μm$^2$. Figure 1(b) shows that there is plenty of room for the cell to be miniaturized further geometrically. At least four flux-quantum states, as shown in Figure 2(b), indicate that the storage loop inductance can be reduced further. The nano-SQUID modulating line, readout micro-SQUID can also be overlaid with the storage loop for space saving. The skewness of the CPR $\Delta\theta \sim \pi/2$ now takes away the lower limit of the inductance in creation of a stable bi-state memory hysteresis. Further miniaturization of the memory cell toward dimensions of under 1×1 μm$^2$ with the current photolithography line-width (0.35 μm) will not be difficult. Other Josephson junction technologies that are capable of inducing a similarly skewed CPR should make it possible to build similarly miniaturized memory cells. However, the dimensions of the 3D nano-SQUID also represent a major advantage in the miniaturization procedure. The junction thickness determines both the critical current and the CPR skewness simultaneously. Therefore, use of this approach to fabricate a scalable superconducting memory will be very promising when the 3D nanobridge junctions can be produced with a uniform thickness. Furthermore, future memory development must be resorted to electronic computer-aided designing (EDA) tools. It would be essential to build a supplemental module to the existing superconducting EDA software that can describe a junction with a skewed CPR.

**Conclusions**

In conclusion, we have demonstrated a miniaturized superconducting memory cell with dimensions of 8×9 μm$^2$ using a 3D nano-SQUID. We observed stable periodic hysteresis between the neighbouring flux quantum states produced by sweeping the bias current to the storage loop. The hysteresis loop size can be tuned further by flux tuning of the critical current of the nano-SQUID for cross-selected memory cell implementation. The CPR can be obtained by measuring the bias current as a function of the total magnetic flux in the storage loop. The measured CPRs of the nano-SQUID and the NBJ are both skewed from that of an ideal Josephson junction and show good agreement with the predictions of theoretical models. Both the critical current and the CPR skewness are linearly correlated and are determined by the nanobridge junction thicknesses. Furthermore, the normalised hysteresis size scales linearly with the CPR skewness, which agrees well with our theoretical prediction. Therefore, the CPR skewness plays a role equivalent to that of the inductance in forming a stable memory hysteresis. An appropriate skewness range from $\pi/4$ to $3\pi/4$ can mitigate the minimal loop inductance requirement. Further miniaturization of our proposed memory cell towards dimensions of ~1×1 μm$^2$ should be technically straightforward. The skewed CPR allows the memory cell to overcome the size limitation due to the inductance and therefore will be an important building block for scalable superconducting memory fabrication.

**Methods**

**Sample Fabrication.**

Device fabrication was performed using the 3D-NBJ fabrication process.[35] First, a Nb film layer with a thickness of 150 nm, which is coloured blue in Figure 1(b), was deposited *via* direct current magnetron sputtering on a 4-in silicon wafer. After photolithography and reactive-ion etching processes, the photo-resist was left on here. A 25-nm-thick SiO$_2$ layer was deposited to produce a ~12-nm-thick SiO$_2$ layer on the sidewall of the first Nb layer. Then, a second Nb film with a thickness of 150 nm, which is coloured yellow in Figure 2(b), was deposited. Using a lift-off process, a 12-nm-wide SiO$_2$ insulator slit was then formed between the two Nb banks. Finally, Nb nanobridges that were ~10-nm-thick and 50-nm-wide were patterned across the insulating gap by electron-beam lithography. Here, we used a 10-nm-thick and 800-nm-wide Nb patch to form a closed storage loop. Devices with both a single NBJ and a 3D nano-SQUID were designed and fabricated from the same 4-in wafer batch. All devices were tested in liquid helium at 4.2 K.

**Electrical Measurement.**

The bias current $I_{wx}$, $I_{wy}$, $I_{rx}$, and $I_{ry}$ as shown in Figure 1(a) are supplied by four independent DC current sources. $I_{wx}$ generates a magnetic flux to the nano-SQUID through an external coil, and $I_{wy}$ is connected to the storage loop directly. During the sweeping of $I_{wy}$, the voltage of the readout SQUID is monitored by a voltage meter and locked to a fixed value by applying feedback current through $I_{rx}$. Therefore, the change in the magnetic flux of the storage loop is given by the change of $I_{rx}$. The mutual inductance between the storage loop and the readout SQUID loop is 1.62 pH.

**CPR and Hysteresis Size**

As shown in Figure 5(a), the current $I_w$ that is injected into the storage loop with a junction will be split into the current $I_j$, which is directed to the junction, and ($I_w-I_j$), which is directed to the right part of the loop. The total flux in the storage loop $\Phi_T$ can then be written as $\Phi_T = (I_w-I_j)L_r - I_j(L_j+L_l) = I_w L_r - I_j L_T$, where $L_j$, $L_r$ and $L_l$ are the inductances of the junction, the right part and the left part of the storage loop, respectively. Here, $L_T = L_j+L_r+L_l$. The junction current $I_j = I_0 f(\theta)$, where $\theta$ is the phase difference across the junction and $f(\theta)$ is the CPR of the junction with the amplitude $I_0$. Additionally, there is a magnetic flux quantisation effect in a superconducting loop given by $\theta/2\pi - \Phi_T/\Phi_0 = n$, where $n=0, \pm 1, \pm 2$ represents the different flux quantum states. The above equations can then be written together as $I_0 L_T f(2\pi\Phi_T/\Phi_0)/\Phi_0 = I_w/I_{w\text{-period}} - \Phi_T/\Phi_0$, where $I_{w\text{-period}} = \Phi_0/L_r$. Therefore, we can obtain the CPR of a junction by measuring $I_w$ as a function of $\Phi_T$, as shown in Figure 5(b) with a skewness $\Delta\theta$ of $\pi/4$.

With a known CPR, we can then also plot $I_w/I_{w\text{-period}} = I_0 L_T f(2\pi\Phi_T/\Phi_0) + \Phi_T/\Phi_0$, which is the sum of a periodic function and a straight line of $\Phi_T/\Phi_0$. As shown in Figure 5(c), $I_w$ was plotted as a function of $\Phi_T$ with $I_0 L_T$ varying from $0.1\Phi_0$ to $0.4\Phi_0$ (from white to blue) for an ideal Josephson effect of $f(\theta)=\sin\theta$. To estimate the hysteresis loop size $\Delta I_w/I_{w\text{-period}}$, $\Delta I_w$ is approximated to be the difference in the $I_w$ values corresponding to the maximal ($\theta=\pi$, $\Phi_T=0.25\Phi_0$) and minimal ($\theta=3\pi/2$, $\Phi_T=0.75\Phi_0$) values of the sine function. We then obtain the relation $\Delta I_w/I_{w\text{-period}} = 2I_0 L_T/\Phi_0 - 0.5$. Therefore, to observe a stable hysteresis, the condition that $I_0 L_T \geqslant 0.25\Phi_0$ must be fulfilled for $\Delta I_w/I_{w\text{-period}} \geqslant 0$. In the skewed CPR case, the phases of the maximal and minimal values in the CPR deviate by $\Delta\theta$. Then, $\Delta I_w/I_{w\text{-period}} = 2I_0 L_T/\Phi_0 + (\Delta\theta/\pi) - 0.5$ is obtained. The jump height between two flux quantum states can then be estimated geometrically by drawing a parallelogram with known height. Therefore, we obtain the relation $\Delta\Phi_T = 2I_0 L_T/(1+\Delta I_w/I_{w\text{-period}})$, as shown in Figure 5(c).

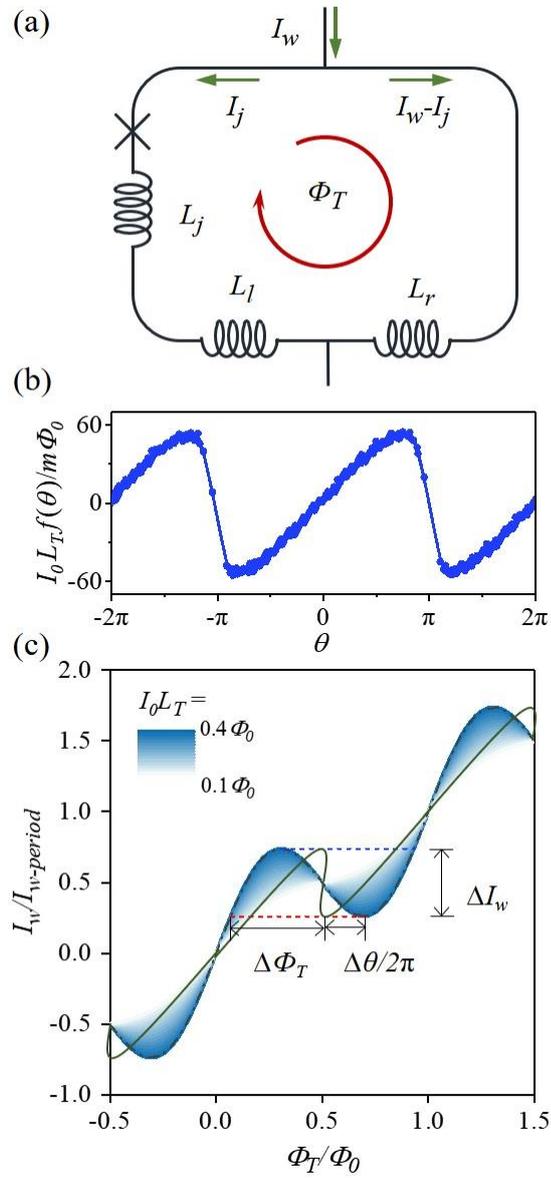

**Figure 5.** (a) Equivalent electrical circuit of a storage loop. (b) Current-phase relationship obtained by measuring $I_w$ as a function of $\Phi_T$. (c) Calculated $I_w$ as a function $\Phi_T$ with a sine CPR and a skewed CPR (green dash-dot line).

## Acknowledgements

This work was supported by the Frontier Science Key Programs of the CAS (Grant No. QYZDY-SSW-JSC033), the Young Investigator program of the CAS (Grant No. 2016217), the Strategic Priority Research program of the CAS (Grant No. XDA18000000) and the National Science Foundation of China (Grant No. 11827805), the National Key R&D Program of China (Grant No. 2017YFF0206105). The authors also appreciate all the Shanghai medical personnel that provide us a safe community from COVID-19 to write this paper.

## Author Contributions

Lei Chen and Zhen Wang planned the research and wrote the paper. Lili Wu and Lei Chen performed the experiments and collected the data. Yue Wang, Denghui Zhang, Yinping Pan, Junwen Zeng and Yihua Wang participated and assisted the experiments. Xiaoyu Liu, Linxian Ma and Wei Peng assisted in the fabrication of devices. Lei Chen, Jie Ren and Zhen Wang performed data analysis. All authors approved the final version of the manuscripts.

## Additional Information

**Competing Interests:** The authors declare no competing interests.

**Table of Content**

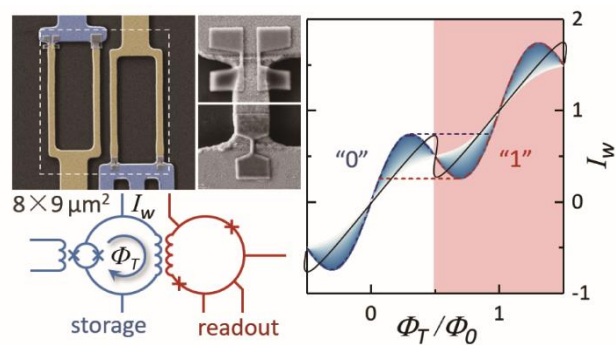